\pdfoutput=1

\documentclass[aps,twocolumn,amsmath,amssymb,preprintnumbers,superscriptaddress,floatfix]{revtex4}
\usepackage[utf8]{inputenc}
\usepackage{newtxtext}
\usepackage[upint]{newtxmath}
\usepackage{microtype}
\usepackage{textcomp}
\usepackage{eucal}
\usepackage{bm}
\usepackage{siunitx}

\usepackage{enumerate}
\usepackage{amsfonts}
\usepackage{amsmath}
\usepackage{amssymb}
\usepackage{color}
\usepackage{soul}

\usepackage{graphicx}

\usepackage[colorlinks,allcolors=blue]{hyperref}
\usepackage[capitalize]{cleveref}

\let\phi=\varphi
\let\epsilon=\varepsilon

\newcommand{\etal}{et al.\ }
\newcommand{\eg}{e.g.\ }

\definecolor{DarkRed}{rgb}{0.80,0,0}

\newcommand{\prlsection}[1]{\textit{#1}.\kern0.05em---\kern0.05em\ignorespaces}

\begin{document}
\title{Spin accumulation induced by a singlet supercurrent}
\author{Morten Amundsen and Jacob Linder}
\affiliation{Center for Quantum Spintronics, Department of Physics, Norwegian \\ University of Science and Technology, NO-7491 Trondheim, Norway}
\begin{abstract}
We show that a supercurrent carried by spinless singlet Cooper pairs can induce a spin accumulation in the normal metal interlayer of a Josephson junction. This phenomenon occurs when a nonequilibrium spin-energy mode is excited in the normal metal, for instance by an applied temperature gradient between ferromagnetic electrodes. Without supercurrent, the spin accumulation vanishes in the Josephson junction. With supercurrent, a spatially antisymmetric spin accumulation is generated that can be measured by tunneling to a polarized detector electrode. We explain the physical origin of the induced spin accumulation by the combined effect of a Doppler shift induced by a flow of singlet Cooper pairs, and the spin-energy mode excited in the normal metal. This effect shows that spin control is possible even with singlet Cooper pairs in conventional superconductors, a finding which could open new perspectives in superconducting spintronics.
\end{abstract}
\maketitle


\prlsection{Introduction}. Using superconductors to achieve interesting spin-dependent quantum effects is the central goal in the growing field of superconducting spintronics \cite{linder_nphys_15, eschrig_rpp_15}. Despite the fact that superconductivity is usually antagonistic to magnetism, a series of experiments have in recent years proven that superconductors can be used to achieve phenomena such as long-ranged and dissipationless spin currents \cite{robinson_science_10, khaire_prl_10}, large thermoelectric effects when combined with spin-polarized barriers \cite{kolenda_prl_16}, spin Hall signals exceeding the normal-state value by three orders of magnitude \cite{wakamura_natmat_15}, and quantum phase batteries \cite{szombeti_nphys_16}. 

A key component of superconducting spintronics has traditionally been to find ways to generate polarized triplet Cooper pairs which can transport spin without resistance. In contrast, conventional superconductors described by Bardeen-Cooper-Schrieffer theory \cite{bardeen_pr_57} are condensates of singlet Cooper pairs. While such condensates support supercurrents of charge, they do not generate supercurrents of spin. It might therefore seem like supercurrents in conventional superconductors do not have much use in spintronics, where the aim is to control and detect spin-polarized signals \cite{zutic_rmp_04}. 

Here, we show that supercurrents carried by singlet Cooper pairs can induce a spin accumulation in a normal metal despite the fact that they have no spin. This phenomenon occurs when a nonequilibrium spin-energy mode is excited in the normal metal. We show that the induced spin accumulation can be understood physically from the combined effect of a Doppler shift induced by the supercurrent and the existence of a spin-energy excitation in the normal metal. The fact that the spin accumulation can be controlled by a superflow of spinless Cooper pairs opens up for a different way in which conventional superconductors can merge with spintronics.

\begin{figure}[t!]
\includegraphics[width=0.5\textwidth]{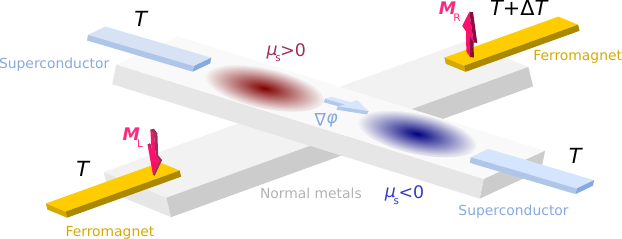}
\caption{(Color online) Two superconducting electrodes (S) are deposited on top of a normal metal film (purple region). Far away from the superconducting electrodes, two antiparallel ferromagnets are in contact with the normal metal film. When a temperature gradient is applied between the ferromagnets, a spin-energy mode is excited throughout the normal metal. In the middle of the normal metal, between the superconductors, there exists no spin accumulation in the absence of a supercurrent. This corresponds to zero phase-gradient across the Josephson junction, $\nabla\phi=0$. When a supercurrent is applied, $\nabla\phi \neq 0$, an antisymmetric spin accumulation is induced in the normal metal between the superconductors. }
\label{fig:model}
\end{figure}

\prlsection{Results}. The proposed setup for measuring this effect is shown in Fig. \ref{fig:model}. Two thin normal metals are stacked on top of each other, creating a four-terminal device. Two ferromagnetic leads with antiparallel magnetizations are attached to opposite terminals, and superconducting leads are attached to the remaining two terminals. When a supercurrent is passed through the superconducting electrodes, a spin accumulation is generated in the normal metal separating them. In the absence of supercurrent, the spin accumulation vanishes in the region between the superconductors. The length of each of the normal metals is assumed to be $3\xi$, where $\xi$ is the coherence length of the superconductors, which then gives the distance between opposite terminals. We assume that the system is in the diffusive limit, with a short mean free path. On the other hand, the spin flip scattering length is assumed to be longer than the size of the system, so that the spin diffusion in the normal metal is negligible. This is achievable, e.g., by using niobium superconductors, which has a coherence length of $\xi = $10-\SI{15}{\nano\meter} in the diffusive limit, and copper normal metals, in which the spin diffusion length at low temperatures can be longer than 100 \si{\nano\meter}, even with a high concentration of impurities~\cite{bass_jpcm_2007,fowler_jmmm_2009}.

The physical mechanism behind this result can then be understood by the following simplified picture. Consider a ferromagnet - normal metal - ferromagnet (FNF) spin valve, with an antiparallel orientation of the magnetization in the ferromagnets. We increase the temperature of the right F by a certain amount $\Delta T$ relative to the temperature $T_0$ of the left F. The tunnelling amplitude of particles at the F-N metal interfaces is higher when their spin is parallel to the magnetization than if it is antiparallel. The former is therefore influenced by a temperature increase in the F reservoir to a greater degree than the latter, leading to a temperature difference between particles of opposite spin. The temperature on the right and left side of the normal metal for spin $j$, $T_R^j$ and $T_L^j$, respectively, are then given as
\begin{align}
T_R^{\uparrow} =& T_0 + \Delta T, \\
T_R^{\downarrow} =& T_0 + (1-P)\Delta T, \\
T_L^{\uparrow} =& T_0 + P\Delta T, \\
T_L^{\downarrow} =& T_0,
\end{align} 
where the polarization $P\in [0,1]$ takes into account the spin dependence of the tunnelling. For $P = 0$, both sides are given the temperature of their respective reservoir, regardless of spin. For $P = 1$, $T_R^{\downarrow}$ and $T_L^{\uparrow}$ are completely insulated from the adjacent interface, and thus equilibrate to the temperature of the reservoir at the opposite end. The temperature distribution throughout the normal metal is simply given by
\begin{align}
T^j(x) = \frac{1}{2}\left(T_R^j + T_L^j\right) + \left(T_R^j - T_L^j\right)\frac{x}{L},
\end{align} 
where $L$ is the distance between the ferromagnets and $x\in\left(-L/2 , L/2\right)$. The temperature difference $T_s$ between spin up and spin down electrons then becomes
\begin{align}
T_s(x) = T^{\uparrow}(x) - T^{\downarrow}(x) = P\Delta T.
\label{eq:ts}
\end{align}
In other words, a spin valve in the antiparallel configuration gives a spatially constant temperature difference between electrons of opposite spin.

When the superconducting leads are added to the spin valve as shown in \cref{fig:model}, the picture is modified. In a superconductor,  any temperature difference between spins of quasiparticles with energies below the superconducting gap will decay with a length scale of the superconducting coherence length, as these particles convert into singlet Cooper pairs. The superconducting correlations induced in the normal metal via the proximity effect therefore has a detrimental effect on $T_s$. The decay is largest near the superconducting leads, where the superconducting correlations are greatest. In addition, heat transfer between the superconducting leads (where both spin species have the same temperature) and the normal metal reduces $T_s$ as well. $T_s$ is therefore expected to have a transversal variation, with a maximum at the center of the spin valve.
\begin{figure}[t]

\includegraphics[width=0.47\textwidth]{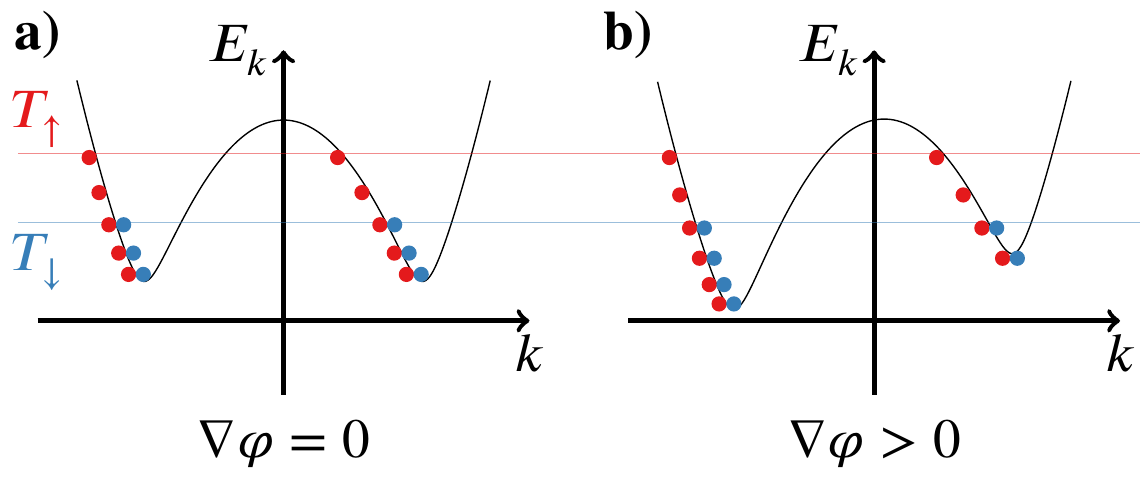}
\caption{Illustration of how a phase gradient in a superconductor leads to a spin imbalance when there is a gradient in the temperature difference between spin up and spin down particles, $T_s = T_{\uparrow} - T_{\downarrow}$. a) The energy band of the superconductor when there is no phase gradient. b) The same energy band when $\nabla\phi > 0$. The resulting Doppler shift leads to a net spin imbalance. We show here for simplicity an extreme example of a case where $\nabla T_s < 0$, in which there are only left moving quasiparticle excitations. }
\label{fig:spinimbalance}
\end{figure}
However, a nonuniform $T_s$ is not enough to generate a spin accumulation. A phase gradient parallel to $\nabla T_s$ is required as well. This is illustrated in \cref{fig:spinimbalance}, which shows the quasiparticle energy band of 
a superconductor with an applied spin temperature gradient. For simplicity, we assume that the temperature on its left side is so low that there are no right-moving quasiparticle excitations. On its right side, a higher temperature is applied, along with a $T_s > 0$, both of which reduce towards the left. When the phase gradient $\nabla\phi$ is zero, in \cref{fig:spinimbalance}a), there is indeed a higher number of spin up quasiparticles excited than spin down. However, there are just as many hole-like excitations as there are electron-like. There is therefore no net spin accumulated. In contrast, when $\nabla\phi > 0$, a Doppler shift of the energy band is created, reducing the gap for momentum $k < 0$, and vice versa, as shown in \cref{fig:spinimbalance}b). This creates an imbalance between spin up and spin down excitations, resulting in a net magnetization.
     
To summarize, the simplified analysis above implies the generation of a spin accumulation in the system shown in \cref{fig:model}. The role of the singlet superconductors is twofold. Firstly, they introduce a transversal variation to an otherwise constant temperature difference between spin up and down particles. Secondly, when a phase gradient is applied parallel to $\nabla T_s$, a net spin imbalance is produced.  To prove this, we have to consider both the superconducting correlations induced in the normal metal due to the proximity effect as well as the non-equilibrium population of quasiparticles caused by the temperature gradient applied across the normal metal. A suitable theoretical framework for this purpose is the Keldysh-Usadel theory for non-equilibrium Green functions \cite{usadel_prl_70, rammer_rmp_86}. In recent years, this formalism has been used to predict several interesting phenomena in superconducting hybrid structures driven out of equilibrium \cite{beckmann_jpcm_2016, bergeret_rmp_2018, rabinovich_prl_2019}. We consider the diffusive regime of transport, where impurity scattering randomizes the momentum of quasiparticles, in which case the Green function matrix in the normal metal can be obtained by solving the Usadel equation,
\begin{align}
D\nabla\cdot\check{g}\nabla\check{g} + i\left[\varepsilon\check{\rho}_4\;,\;\check{g}\right] = 0,
\label{eq:usadel}
\end{align}
where $D$ is the diffusion constant and $\varepsilon$ is the quasiparticle energy. The Green function matrix has the structure
\begin{align}
\check{g} = \begin{pmatrix} \hat{g}^R & \hat{g}^K \\ 0 & \hat{g}^A \end{pmatrix},
\label{eq:gf}
\end{align}
in Keldysh space, where $\hat{g}^X$ are $4\times4$ matrices in particle-hole and spin space. Furthermore, we have $\check{\rho}_4 = \text{diag}\left(\hat{\rho}_4\;,\;\hat{\rho}_4\right)$, with $\hat{\rho}_4 = \text{diag}\left(+1,+1,-1,-1\right)$. The retarded and advanced Green functions, $\hat{g}^R$ and $\hat{g}^A$, determine the band structure of the system, and these components satisfy an equation which is identical in form to \cref{eq:usadel}. The quasiparticle excitations are determined by the Keldysh Green function, $\hat{g}^K$. Without loss of generality, this matrix can be parametrized has $\hat{g}^K = \hat{g}^R\hat{h} - \hat{h}\hat{g}^A$, where $\hat{h}$ is a distribution function. Its matrix structure in particle-hole and spin space can be further parametrized as
\begin{align}
\hat{h} = \sum_{n} h_n\hat{\rho}_n,
\end{align}
where $\hat{\rho}_0 = \hat{I}$, $\hat{\rho}_j = \hat{\sigma}_j$, and $\hat{\rho}_{4+j} = \hat{\rho}_4\hat{\rho}_j$  for $j\in\left\{1,2,3\right\}$. The matrix $\hat{I}$ is the identity, and $\hat{\sigma}_j = \text{diag}\left(\sigma_j , \sigma_j^*\right)$ for Pauli matrix $\sigma_j$. In the following, we assume that both ferromagnets are aligned in the $z$ direction, in which case the only relevant distribution functions become $h_0$, $h_3$, $h_4$ and $h_7$. Insertion into \cref{eq:usadel} gives,
\begin{align}
a_{mn}\nabla^2 h_n + \bm{b}_{mn}\cdot\nabla h_n = 0,
\label{eq:dist}
\end{align}
where $a_{mn} = D\text{Tr}\left[\hat{\rho}_m\hat{\rho}_n - \hat{\rho}_m\hat{g}^R\hat{\rho}_n\hat{g}^A\right]/4$, and $\bm{b}_{mn} = \nabla a_{mn} + D\text{Tr}\left[\hat{\rho}_n\hat{\rho}_m\hat{g}^R\nabla\hat{g}^R - \hat{\rho}_m\hat{\rho}_n\hat{g}^A\nabla\hat{g}^A\right]/4$. The function $h_0$ is the energy mode, and gives the temperature distribution of the system, with $h_0 = \tanh \frac{\varepsilon}{2k_B T}$, where $k_B$ is the Boltzmann constant, being the only nonzero component of $\hat{h}$ in equilibrium. $h_3$ is a spin-energy mode, and expresses an effective temperature difference between spin up and down quasiparticles. The charge mode $h_4$ gives the quasiparticle charge distribution in the system, and the spin mode $h_7$ gives the spin accumulation, through the relation
\begin{align}
\mu_s(\bm{r}) = 4\mu_0 \int d\varepsilon\; h_7(\varepsilon,\bm{r})\nu(\varepsilon,\bm{r}). 
\label{eq:magnetization}
\end{align}
In \cref{eq:magnetization}, we have neglected any triplet superconducting correlations, as is the case in our system, which would otherwise also give a contribution. Furthermore, $\nu(\varepsilon,\bm{r})$ is the local density of states, and $\mu_0 = g\mu_B\nu_0/8$, where $g$ is the Landé $g$-factor, $\mu_B$ is the Bohr magneton and $\nu_0$ is the density of states of the normal metal, at the Fermi level.

\begin{figure}[t]
\includegraphics[width=0.5\textwidth]{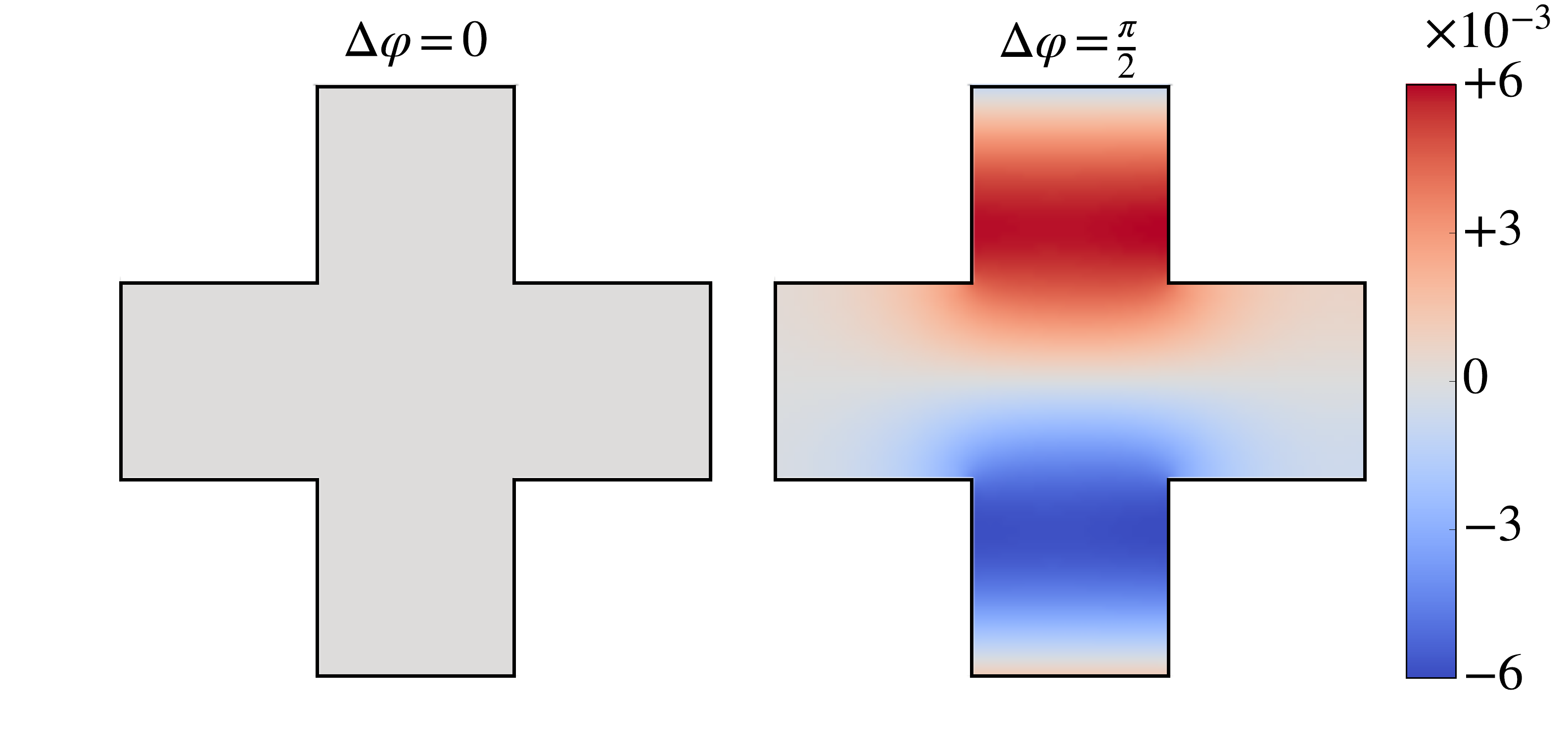}
\caption{Numerical simulations of the spin accumulation, scaled by $\mu_0$, in the presence of temperature gradient. The superconducting terminals (top and bottom), and the left ferromagnet have a temperature of $T_l = 0.1T_c$, whereas the right ferromagnet has a temperature of $T_h = 0.5T_c$. No magnetization is induced when $\Delta\phi = 0$.}
\label{fig:magnetization}
\end{figure}

To describe the interfaces to the reservoirs, we use a generalization of the Kupriyanov-Lukichev tunnelling boundary conditions, which take spin polarization into account~\cite{kupriyanov_jetp_1988,eschrig_njp_2015}, 
\begin{align}
\zeta\hat{n}\cdot\check{g}\nabla\check{g} = \left[\check{g}',\check{g}\right] + \zeta_{mr}\left[\left\{\check{\sigma}_3,\check{g}'\right\},\check{g}\right] + \zeta_1\left[\check{\sigma}_3\check{g}'\check{\sigma}_3,\check{g}\right],
\end{align}
where $\hat{n}$ is the interface normal, $\zeta$ expresses the interface resistance, and $\check{g}'$ is the reservoir Green function. The parameters $\zeta_{mr} =P/\left(1+\sqrt{1-P^2}\right)$ and $\zeta_1 = \left(1 - \sqrt{1-P^2}\right)/\left(1 + \sqrt{1-P^2}\right)$ give the spin filtering at the interface, for a given polarization $P$. For interfaces to the ferromagnets, we set $\zeta = 3$ and $P = 0.6$, whereas for the superconductors, we set $\zeta = 1$ and $P = 0$. To generate a temperature gradient in the normal metal, we set the temperature in the left ferromagnetic reservoir, as well as in the two superconducting leads to be $T_l = 0.1T_c$, and the temperature in the right ferromagnetic reservoir to be $T_h = 0.5T_c$. The retarded and advanced components of \cref{eq:usadel}, and subsequently, \cref{eq:dist}, are solved using the finite element method~\cite{amundsen_scirep_2016}, and the resulting magnetization is computed using \cref{eq:magnetization}. The results are shown in \cref{fig:magnetization}. It is seen that when the phase difference $\Delta\phi$ is zero, no magnetization is induced in the normal metal. In stark contrast, an antisymmetric magnetization appears when $\Delta\phi = \pi/2$. Thus, a supercurrent carried by spinless Cooper pairs induces a magnetization.

A magnetization can also be generated due to the presence of the ferromagnets, which in proximity to a superconductor can produce triplet superconducting correlations~\cite{buzdin_rmp_2005}. Another source of triplet correlations are the spin filtering at the interfaces, which would polarize the supercurrent if it detours via the ferromagnets on its way from one superconducting lead to the other. However, for the present geometry, the ferromagnets are located sufficiently far away from the superconducting leads that these mechanisms can be disregarded. In other words, the triplet superconducting correlations are completely negligible in this system, and the magnetization is induced solely by the interaction between the singlet (spin-0) Cooper pairs and the nonequilibrium temperature distribution. 

\begin{figure}[t]
\includegraphics[width=0.5\textwidth]{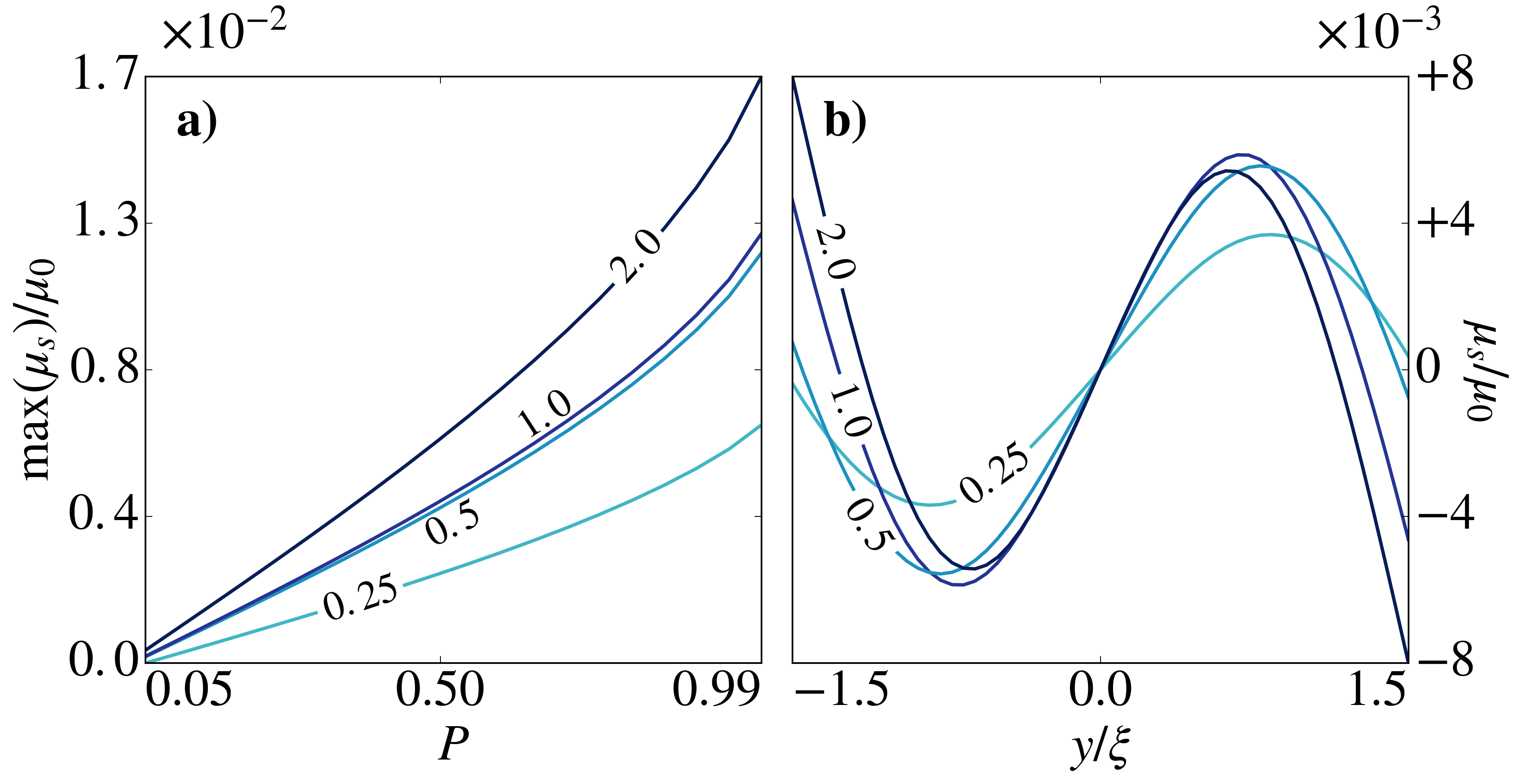}
\caption{The effect of varying the temperature in the right ferromagnetic reservoir. a) shows the maximum spin accumulation as a function of the interface polarization $P$, and b) shows the distribution of the spin accumulation along a coordinate $y$ moving in a straight line between the superconductors. The annotations denote different $T_h/T_c$, and $\text{max}(\mu_s)$ is the maximum spin accumulation located along $y$.}
\label{fig:pt}
\end{figure}

It is interesting to investigate how the induced magnetization depends on the system parameters. In \cref{fig:pt}a) we show the maximum spin accumulation as a function of the interface polarization $P$, for a variety of different temperature gradients. Polarizations up to 90\% can be obtained \eg by replacing the ferromagnetic reservoirs in Fig. \ref{fig:model} with normal reservoirs that couple to the central normal metal via a ferromagnetic insulator such as EuS \cite{muller_prl_1972}.  It is seen that the magnetization increases with $P$, and that this increase is steeper for higher $T_h$. This result is reasonable, as both parameters combined generate a spin temperature difference $T_s$, in correspondence with \cref{eq:ts}. \cref{fig:pt}b) shows the distribution of the spin accumulation along a straight line between the superconducting leads. For a low temperature difference $T_h - T_l$, the largest spin accumulation takes place about half way between the center of the system and the superconductors. However, as $T_h$ increases these maxima are eventually overtaken by a larger spin accumulation at the superconductor interfaces. This is likely because of the increasing temperature gradient between the right ferromagnet and the superconductors, which leads to an increasing heat exchange between the two. Since the temperature in the latter is spin independent, this serves to mollify the spin temperature difference $T_s$ near the superconductors, and thus increase the gradient in $h_3$. This, in turn, leads to a higher spin accumulation when a phase gradient is applied. We note, however, that these results are obtained while assuming the superconductors act as temperature reservoirs. A continued increase in $T_h$ will likely invalidate this assumption, and lead to a saturation of the induced spin accumulation.  We also note that a spin-heat accumulation, described by a finite $h_3$ and $T_s$, should in general also occur close to the interface on the ferromagnetic side~\cite{dejene_nphys_2013}.

Finally, we remark that it is also possible to generate a spin dependent temperature difference by applying a voltage bias between the ferromagnets, rather than a temperature gradient. In this case, the largest average $T_s$ in the system would be obtained for a parallel alignment of the ferromagnets. A phase gradient between the superconducting leads will then produce a spin accumulation by the same mechanism as previously described. However, in addition to providing a $T_s$, the injected quasiparticles lead to a spin imbalance, and thus directly contribute to the spin accumulation. This spin accumulation is independent of the phase gradient, and will likely dominate any measurement.


\prlsection{Conclusion}. We have shown that a supercurrent carried by spinless Cooper pairs can induce a spin accumulation in a normal metal. This is possible when a spin-energy distribution mode is excited in the normal metal out of equilibrium, which allows a spin accumulation to arise due to the Doppler shift caused by the supercurrent in the quasiparticle energies. Our finding shows that spin control is possible even with singlet Cooper pairs in conventional superconductors, which could open interesting avenues for further research in superconducting spintronics.


\vspace{3ex}
\begin{acknowledgments}
M.A. and J.L. were supported by the Faculty of Sciences, NTNU, and the Centres of Excellence funding scheme from the Research Council of Norway, grant 262633 ``\emph{QuSpin}''. We thank J. A. Ouassou for useful discussions.
\end{acknowledgments}



\end{document}